\documentclass[twocolumn]{aastex63}
\usepackage[normalem]{ulem}  
\usepackage[T1]{fontenc}
\usepackage{ae,aecompl}
\usepackage{graphicx}
\usepackage{amsmath}
\usepackage{amssymb}
\usepackage{supertabular}
\usepackage{hyperref}
\usepackage{xcolor}
\usepackage{bm}
\usepackage{academicons}

\def\bea{\begin{eqnarray}} \def\eea{\end{eqnarray}}
\def\fm3{\;\text{fm}^{-3}}
\def\mev{\;\text{MeV}}
\newcommand{\Msun}{\,{M}_{\odot}}

\newcommand{\nicer}{{\it NICER}}

\newcommand{\ie}{i.e.,~}
\newcommand{\eg}{e.g.,~}

\begin{document}

\title{A Bayesian inference of relativistic mean-field model for neutron star matter from observation of NICER and GW170817/AT2017gfo}

\author[0000-0001-9189-860X]{Zhenyu Zhu}
\affiliation{Tsung-Dao Lee Institute, Shanghai Jiao Tong University, Shanghai 201210, China; zhenyu.zhu@sjtu.edu.cn}

\author[0000-0001-9849-3656]{Ang Li}
\affiliation{Department of Astronomy, Xiamen University, Xiamen, Fujian 361005, China; liang@xmu.edu.cn; tongliu@xmu.edu.cn}

\author[0000-0001-8678-6291]{Tong liu}
\affiliation{Department of Astronomy, Xiamen University, Xiamen, Fujian 361005, China; liang@xmu.edu.cn; tongliu@xmu.edu.cn}

\date{\today}

\begin{abstract} 
  The observations of optical and near-infrared counterparts of binary neutron star mergers 
  not only enrich our knowledge about the abundance of heavy elements in the Universe, or help reveal the remnant object just after the merger as generally known, but also can effectively constrain dense nuclear matter properties and the equation of state (EOS) in the interior of the merging stars.
  Following the relativistic mean-field description of nuclear matter, we perform the Bayesian inference of the EOS and the nuclear matter properties using the first multi-messenger event GW170817/AT2017gfo, together with the NICER mass-radius measurements of pulsars.
  The kilonova is described by a radiation-transfer model with the dynamical ejecta,
  and light curves connect with the EOS through the quasi-universal relations between the ejecta properties (the ejected mass, velocity, opacity or electron fraction) and binary parameters (the mass ratio and reduced tidal deformability). 
  It is found that the posterior distributions of the reduced tidal deformability from the AT2017gfo analysis display a bimodal structure, with the first peak enhanced by the GW170817 data, leading to slightly softened posterior EOSs, while the second peak cannot be achieved by a nuclear EOS with saturation properties in their empirical ranges. 
  The inclusion of NICER data results in stiffened EOS posterior because of the massive pulsar PSR J0740+6620.
  We give results at nuclear saturation density for the nuclear incompressibility, the symmetry energy and its slope, as well as the nucleon effective mass, from our analysis of the observational data.
\end{abstract}


\section{Introduction}
\label{sec:intro}
The detection of gravitational waves (GWs) and light from the binary neutron star merger GW170817 marked the first milestone of multimessenger astronomy~\citep{Abbott2017_etal}. The GW signals from coalescing binary neutron stars have been widely used to provide critical insights into
the nature of dense nuclear matter and the equation of state (EOS; i.e., the pressure-density relation) of neutron stars~\citep{Abbott2018b}. 
The electromagnetic counterparts of GW sources provide another way of studying the EOS. 
In particular, the transient optical/infrared/UV event (AT2017gfo) was detected several hours after the merger time of GW170817~\citep{Andreoni2017, Arcavi2017, Coulter2017, Cowperthwaite2017, Diaz2017, Drout2017, Evans2017, Hu2017, Kasliwal2017, Lipunov2017, Pian2017, Pozanenko2018, Shappee2017, Smartt2017, Tanvir2017, Troja2017, Utsumi2017, Valenti2017}, the luminosity, spectrum and light curve of which are consistent with the prediction of the kilonova model, which attribute its emission to the $r$-process nucleosynthesis of the ejected neutron-rich matter from the merger. 
The mass, velocity and electron fraction of the ejecta are key parameters for understanding the observations of AT2017gfo~\citep[e.g.,][]{Metzger2017,Perego2017,Yu2018,Ren2019,Qi2022},
and closely related to the binary parameters (like the mass ratio, the radius) and the EOS~\citep[e.g.,][]{ShibataRev19}. 

Stiffness or softness of the EOS implies larger or smaller stellar radius and orbital separation at merger. A softer EOS and smaller radius results in a more violent collision and more efficient
shock heating, which can eject more material with higher velocity and high temperature. The ejected matter with high temperature may trigger the weak interaction and neutrino emission, and further vary the electron fraction of ejecta. Therefore, the EOS affects the input quantities of the kilonova light curve model, and it is interesting and important to infer the EOS from both the GW and kilonova data.

Merger simulations have revealed some quasi-universal relations of ejecta properties and binary parameters (mass ratio and reduced tidal deformability)~\citep{Nedora2021}.
The EOS constraints from kilonova observation have also been investigated \citep[e.g.,][]{Margalit2017,  Radice2017b, Coughlin19, Breschi2021, Holmbeck2022}.
A group of EOSs from different nuclear many-body frameworks, or the parameterizations of EOS such as piecewise polytropes~\citep{Most2018,  De2018, Ecker2022}, or spectral parametrization~\citep{Lindblom2010, Koliogiannis2019} were usually adopted, allowing the study only on the pressure-versus-density function, but not on the physical properties of nuclear matter.

In this work, we perform one of the first studies to connect nuclear matter microscopic parameters to the AT2017gfo data~\citep{Villar2017} of the GW170817 binary neutron star merger.
The kilonova is described by a radiation-transfer model depending on which we reproduce important properties of AT2017gfo light curves and explore the underlying phase state of nuclear matter and the EOS. 
Nuclear matter and the EOS are described by the relativistic mean-field (RMF) model, 
which encodes a great amount of nuclear physics in a handful of model parameters.
By construction, the RMF effective interactions can facilitate easy incorporation of various nuclear EOS constraints at the nuclear saturation density $n_0$ and moderate values of the isospin asymmetry. 
In combination with the GW observations of tidal deformability~\citep{Abbott2019c} by LIGO/Virgo and the mass and radius measurements of PSR J0030+0451 and PSR J0740+6620~\citep{MCMiller2019b,
Riley2019, Miller2021, Riley2021} by the NASA Neutron Star Interior Composition ExploreR (NICER) mission, the inference will be performed directly on key properties like the nuclear incompressibility and the symmetry energy as well as the single particle nucleon effective mass in medium, that can be confronted with laboratory studies on nuclear structure and reactions.
We do not consider the nonnucleon degree of freedom possibly present in heavy neutron stars since the data we utilize here are mostly from typical stars around or below $1.4\Msun$ and our main interest of the present study is the EOS parameters around the saturation density $n_0$.
Because the stellar radius is controlled mainly by the density dependence of the nuclear symmetry energy around $n_0$~\citep{Lattimer2000}, below we report also the most preferred radius and tidal deformability (scaling as the fifth power of the radius) for typical $1.4\Msun$ stars based on our analysis.
See e.g., \citet{Miao2020, Li2021, Li2021b, Sun2022, Miao2022, Miao2022b} for analysis incorporating strangeness phase transitions in neutron star matter.

The paper is organized as follows.
In Sec.~\ref{sec:model}, we will introduce the models of EOS and kilonova that employed in our analyses. 
In Sec.~\ref{sec:analyses}, we recall the Bayesian formulation and describe the parameters, priors and likelihood functions in our analyses. 
In Sec.~\ref{sec:results}, we present our results and
discussions. Finally, we conclude in Sec.~\ref{sec:sum}.
\section{Models of neutron star EOS and kilonova}
\label{sec:model}
In this section, we will review the adopted models of neutron star EOS and kilonova, including a detailed description of the relations between the kilonova observations and the EOS as well as the stellar properties.

\subsection{neutron star EOS}
\label{sec:eos}
The only physics that spherically-symmetric neutron stars in hydrostatic equilibrium are sensitive to is the EOS of (neutron-rich) nuclear matter, in the simple case of no strangeness phase transition~\citep{Li2020c}.
In principle, it can be determined by the strong interaction, from solving the first principle QCD. Nevertheless, the complexity of nonperturbative strong interaction makes it difficult to do theoretically, and hence parameterization are widely used to describe the EOS in the analyses of observational data. 
Presently, the RMF nuclear many-body model is employed in our analyses.

The RMF model starts from a many-body Lagrangian for describing the nucleon-nucleon interactions, which are mediated by scalar ($\sigma$), isoscalar–vector ($\omega$) and isovector–vector ($\rho$) mesons~\citep[see e.g.,][]{Li2008, Zhu2018, Zhu2019, Traversi2020},
\begin{eqnarray}
  \label{eq:lagrangian}
\mathcal{L}& = & \overline{\psi}\left(i\gamma_\mu \partial^\mu - M_N + g_\sigma \sigma - g_{\omega}\omega\gamma^0 - g_{\rho}\rho\tau_{3}\gamma^0\right)\psi  \nonumber \\
           && -\frac{1}{2}(\nabla\sigma)^2 - \frac{1}{2}m_\sigma^2 \sigma^2 - \frac{1}{3} g_2\sigma^3 - \frac{1}{4}g_3\sigma^4 \nonumber \\
           && + \frac{1}{2}(\nabla\omega)^2 + \frac{1}{2}m_\omega^2\omega^2 + \frac{1}{2}g_{\omega}^2\omega^2 \Lambda_v g_{\rho}^2\rho^2 \nonumber \\
           & & + \frac{1}{2}(\nabla\rho)^2 + \frac{1}{2}m_\rho^2\rho^2\ ,
\end{eqnarray}
where $g_\sigma$, $g_{\omega}$ and $g_{\rho}$ are the nucleon coupling
constants for $\sigma$, $\omega$ and $\rho$ mesons. 
We also include the nonlinear $\sigma$ self-interactions with two parameters $g_2$ and
$g_3$, and the $\omega$-$\rho$ coupling with parameter $\Lambda_v$. 
These six meson coupling parameters can be obtained by fitting the empirical data at the nuclear saturation density $n_0$ (see below in Table \ref{tb:prior}).

The equation of motion for each meson can be generated by the Euler-Lagrangian equation from the Lagrangian and applying the mean-field approximation:
\begin{eqnarray}
  \label{eq:eqs_mot1}
  m_\sigma^2 \sigma + g_2 \sigma^2 + g_3 \sigma^3 & = & g_\sigma n_{\rm S}\ , \\
  \label{eq:eqs_mot2}
  (m_\omega^2 + \Lambda_v g_\omega^2 g_\rho^2 \rho^2) \omega & = & g_\omega (n_{\rm p} + n_{\rm n})\ , \\
  \label{eq:eqs_mot3}
  (m_\rho^2 + \Lambda_v g_\omega^2 g_\rho^2 \omega^2) \rho & = & g_\rho (n_{\rm p} - n_{\rm n})\ .
\end{eqnarray}
where
\begin{eqnarray}
  n_s = \sum_{i=n,p}\frac{1}{\pi^2} \int_0^{p_{\rm F}} \frac{M_{\rm N}^\ast}{\sqrt{M_{\rm N}^{\ast 2}
  + p_{\rm F}^2}} p_{\rm F}^2 dp_{\rm F}
\end{eqnarray}
is the scalar density, the $p_{\rm F}$ denotes the fermi momentum, and
$M_{\rm N}^\ast = M_{\rm N} - g_\sigma \sigma$ is the effective
mass. The number density of proton and neutron are represented by
$n_{\rm p}$ and $n_{\rm n}$, respectively. 
After solving these equations of motion, the energy density and pressure of nuclear matter can be computed by:
\begin{eqnarray}
  \label{eq:ener}
  e & = & \sum_{i=n,p} e_{\rm kin}^i + \frac{1}{2} m_{\sigma}^2 \sigma^2 + \frac{1}{3} g_2 \sigma^3 + \frac{1}{4}g_3 \sigma^4 \nonumber \\
    & & - \frac{1}{2} m\omega^2 \omega^2 - \frac{1}{2} m_\rho^2 \rho^2 - \frac{1}{2} \Lambda_v (g_\omega g_\rho \omega \rho)^2 \nonumber \\
    & & + g_\omega \omega (n_{\rm n} + n_{\rm p}) + g_\rho \rho (n_{\rm p} - n_{\rm n})\ , \\
  \label{eq:press}
  p & = & \sum_{i=n,p} p_{\rm kin}^i - \frac{1}{2} m_{\sigma}^2 \sigma^2 - \frac{1}{3} g_2 \sigma^3 - \frac{1}{4}g_3 \sigma^4 \nonumber \\
  & & + \frac{1}{2} m_\omega^2 \omega^2 + \frac{1}{2} m_\rho^2 \rho^2 + \frac{1}{2} \Lambda_v (g_\omega g_\rho \omega \rho)^2\ .
\end{eqnarray}
To study the structure of neutron stars, we have to calculate the composition and EOS of cold, neutrino-free, catalyzed matter. We require that the neutron star contains charge-neutral matter consisting of neutrons, protons, and leptons ($e^-$, $\mu^-$) in beta equilibrium.
Additionally, since we are looking at neutron stars after neutrinos have escaped, we set the neutrino chemical potentials equal to zero.
Also, we use ultrarelativistic and nonrelativistic approximations for the electrons and muons, respectively, and their contributions to the energy and pressure are merely added to the Eqs.~(\ref{eq:ener})--(\ref{eq:press}). Consequently, the energy density and pressure of neutron star matter are simply the functions of nucleon number density.

For completeness, we also write down the expressions of the symmetry energy
$J_0$, incompressibility $K_0$ and symmetry energy slope $L_0$ at the
saturation density in symmetric nuclear matter
\begin{eqnarray}
  \label{eq:esym}
  J_0 & = & \frac{p_{\rm F}^2}{6E_{\rm F}} + \frac{g_\rho^2}{2[m_\rho^2 + \Lambda_v (g_\omega g_\rho \omega)^2]} (n_{\rm p} + n_{\rm n})\ , \\
  \label{eq:k0}
  K_0 & = & \frac{3p_{\rm F}^2}{E_{\rm F}} + \frac{3M_{\rm N}^\ast p_{\rm F}}{E_{\rm F}} \frac{dM_{\rm N}^\ast}{dp_{\rm F}} + \frac{9g_\omega^2}{m_\omega^2 + \Lambda_v (g_\omega g_\rho \rho)^2}n_0\ , \\
  \label{eq:l0}
  L_0 & = & 3J_0 + \frac{1}{2}\left(\frac{3\pi^2}{2}n_0 \right)^{2/3} \frac{1}{E_{\rm F}} \times \nonumber \\
  & & \left(\frac{g_\omega^2}{m_\omega^2 + \Lambda_v (g_\omega g_\rho \rho)^2}\frac{n_0}{E_{\rm F}}
            - \frac{K_0}{9E_{\rm F}} - \frac{1}{3} \right) \nonumber \\
        & & - \left(\frac{3g_\rho^2}{m_\rho^2 + \Lambda_v (g_\omega g_\rho \omega)^2}\right)^2 \frac{g_\omega^3 \Lambda_v \omega n_0^2}{m_\omega^2 + \Lambda_v (g_\omega g_\rho \rho)^2}\ . \nonumber \\
\end{eqnarray}

To recap, we have six nuclear matter properties: The saturation density $n_0$, energy per
baryon $E/A$, $J_0$, $K_0$, $L_0$ and effective mass $M_{\rm N}^\ast$, to be reproduced to fitting the six model parameters, $g_\sigma$, $g_\omega$ $g_\rho$, $g_2$, $g_3$ and $\Lambda_v$. 
Once the saturation properties of nuclear matter are chosen in their empirical ranges, the six model parameters can be uniquely determined  (see Appendix \ref{A1} for more details) for the calculations of neutron stars. 
In our following analysis, we will directly specify these six saturation properties, rather than
the model parameters, to denote the EOS. 

\subsection{Kilonovae}
\label{sec:kn}
In the present work, we employed a radiation transfer model~\citep[see e.g.,][for more details]{Metzger2017, Yu2018, Ren2019, Qi2022} to calculate the light curves of the kilonova. The emission luminosity is computed by solving the energy conservation equations of ejecta, where the heating of $r$-process nucleosynthesis and the cooling of adiabatic expansion are taken into account. Additionally, the source of kilonova is treated as a blackbody and the spectra are given by the blackbody emission.

The ejecta during and after the binary neutron star merger mainly consists of two components, \ie dynamical ejecta and wind-driven ejecta. The dynamical ejection is driven by the tidal forces during the inspiral and shock heating during the coalescence~\citep{Bovard2017, Radice2018a, ShibataRev19}. The tidal forces eject the material primarily along the direction of the equator with relatively low temperature and low electron fraction (smaller than $0.1$-$0.2$).
Meanwhile, the shock isotropically ejects and heats the material to a high temperature where the weak interaction can be triggered so that the electron fraction increases. Therefore, The ejecta driven by shock heating has a high electron fraction ($Y_{\rm e}$ > 0.25) and distributes evenly along the inclination $\theta$. 
In addition to the dynamic ejecta, the neutrino emissions from the remnant before collapsing to black hole as well as the viscosity could further drive more ejecta (the so-called wind-driven ejecta) from the disc surrounding the remnant, which naturally more subject to e.g., the lifetime of the remnant neutron star. For the present study, we do not consider the wind-driven ejecta when connecting the AT2017gfo observational data with the underlying EOS.

The ejected matter with a low electron fraction that mainly concentrates on the orbital plane will undergo a full $r$-process nucleosynthesis and produce a large amount of lanthanide elements. The high opacity result from the lanthanide elements leads the ejecta on the orbital plane to be the ``red'' component. On the other hand, the ejected material along the polar direction is primarily contributed by the shock heating, and will only experience a partial $r$-process nucleosynthesis, whose lanthanide synthesis is suppressed. consequently, the polar ejecta has a relatively lower opacity and is called ``blue'' component.

The multi-wavelength light curves of AT2017gfo indicate that it cannot be
explained by the models with only one single set of parameters, if
only the power of $r$-process nuclei is taken into
account~\citep{Villar2017}. Therefore, we involved both the ``red''
and ``blue'' component in our model by implementing a
$\theta$-dependent opacity. The ejected material is approximated as
being homologously expanding, and the shell structure is formed
accordingly. Each shell can further be decomposed into two patches
with different opacity, and the interface of these two patches is set
to be $\theta=\pi/4$. Therefore, the opacity can be described by a
step function of inclination angle $\theta$:
\begin{eqnarray}
  \label{eq:kappa}
  \kappa = \left\{
  \begin{matrix}
    \kappa_{\rm low}, & \ \ \ \theta \le \pi/4\ ; \\
    \kappa_{\rm high}, & \ \ \ \theta>\pi/4\ , \\
  \end{matrix}
  \right.
\end{eqnarray}
where opacity is denoted by $\kappa$. $\kappa_{\rm low}$ and
$\kappa_{\rm high}$ are constants and correspond to the ``blue'' and
``red'' component of the kilonova. 

Because of the isotropic distribution of mass, the density is merely 
the function of radial coordinate $r$. This distribution is typically described by a power-law~\citep[see][]{Nagakura2014}, and the density distribution function of radius can be written as:
\begin{eqnarray}
  \label{eq:dens_dist}
  \rho_{\rm ej}(R) = \frac{M_{\rm ej}}{4\pi} (3-\delta) \frac{R^{-\delta}}{R_{\rm max}^{3-\delta}
  - R_{\rm min}^{3-\delta}} \ ,
\end{eqnarray}
where the total mass, the maximal and minimal radius of the ejecta is
denoted by $M_{\rm ej}$, $R_{\rm max}$, and $R_{\rm min}$, respectively. The shell
with the maximum and minimal radius also represents the maximum and minimum
velocity shell through $R_{\rm max}=v_{\rm max}t$ and $R_{\rm min}=v_{\rm min}t$. The index $\delta$ is a constant
between $1$ and $3$. With this distribution function, the mass of each
shell can be calculated by integrating over the radius. 

The emission luminosity can be obtained by solving the equation of
energy conservation:
\begin{eqnarray}
  \label{eq:e_con}
  \frac{dE_{_{i,j}}}{dt} = m_{_{i,j}} \dot{q}_r \eta_{\rm th} - \frac{E_{_{i,j}}}{R_i}\frac{dR_i}{dt} - L_{_{i,j}}\ ,
\end{eqnarray}
where $i, j$ denotes the index of patches (indicating that the patch locates
at the $i$th shell and $j$th inclination angular spacing), and
$m_{_{i,j}}$ represents the mass of the patch. Because the opacity is
a step function and only two value is available in our computation, 
the number of patches for each shell is 2 ($j=1,2$). The first term on the right-hand side of this equation represents the heating of 
$r$-process nucleosynthesis. The $\dot{q}_r$ denotes the radioactive 
power per unit mass and $\eta_{\rm th}$ denotes the thermalization
efficiency. They can be written as~\citep{Korobkin2012, Barnes2016}:
\begin{eqnarray}
  \label{eq:qdot_r}
  \dot{q}_r & = & 4 \times 10^{18} \left[\frac{1}{2} - \frac{1}{\pi}\arctan\left(\frac{t-t_0}{\sigma}\right)  \right]^{1.3}
  \text{erg}\ \text{s}^{-1}\ \text{g}^{-1}\ , \nonumber \\
  \eta_{\rm th} & = & 0.36\left[\exp(-0.56t_{\rm day}) + \frac{\ln(1+0.34t_{\rm day}^{0.74})}{0.34t_{\rm day}^{0.74}} \right]\ ,
\end{eqnarray}
where $t_0=1.3$ s, $\sigma=0.11$ s, and $t_{\rm day}=t/1~\rm day$. The second term represents the adiabatic cooling of the
ejecta, and it can be simplified by using the relation $R_i = v_it$
as $-E_{_{i,j}} / t$. The last term represents the energy that is carried
out by emission, or the luminosity. It can be estimated by:
\begin{eqnarray}
  \label{eq:lum}
  L_{_{i,j}} = \frac{E_{_{i,j}}}{\max[t_{\rm d}^{i,j}, t_{\rm lc}^i]}\ ,
\end{eqnarray}
where the light-crossing time $t_{\rm lc}^{i,j} = R_i / c$, and the
photon diffusion time scale:
\begin{eqnarray}
  \label{eq:t_diff}
  t_{\rm d}^i \approx \frac{3\kappa^{j}}{\Delta \Omega R_i c} m_{\rm ex}^{i,j}\ .
\end{eqnarray}
The diffusion time scale depends on the opacity of the ejecta
$\kappa^j$, which is $\theta$-dependent in our model. The
$m_{\rm ex}^{i,j}$ denotes the exterior mass of the patch, which sums
the mass of the exterior of the $i$th shell for the $j$th patch.

By solving Eq.~(\ref{eq:e_con}) and summing the luminosity of each
shells for a specific patch and at a specific time step, we obtain the bolometric luminosity as a function of time:
\begin{eqnarray}
  \label{eq:bol_lum}
  L_{\rm bol}^{j} = \sum_{i} L_{_{i,j}}\ .
\end{eqnarray}
Additionally, the blackbody spectrum is assumed for the emission and
the effective temperature can be calculated through this bolometric
luminosity:
\begin{eqnarray}
  \label{eq:tem_eff}
  T_{\rm eff}^{j} = \left(\frac{L_{\rm bol}^{j}}{\sigma_{\rm SB}\Delta \Omega R_{\rm ph}^2}\right)^{1/4}\ ,
\end{eqnarray}
where $\sigma_{\rm SB}$ is the Stephan-Boltzmann constant and
$R_{\rm ph}$ is the radius of the photosphere, which is defined as the
radius where the exterior optical depth
$\tau_{R_{\rm max} - R_{\rm ph}}$ is unitary. This radius can be
calculated analytically with the density distribution
Eq.~(\ref{eq:dens_dist}) as
\begin{eqnarray}
  \label{eq:r_ph}
  R_{\rm ph}^\tau = \left[R_{\rm max}^{1-\delta} - \frac{4\pi}{M_{\rm ej} \kappa}
                \frac{1-\delta}{3-\delta} (R_{\rm max}^{3-\delta}
              - R_{\rm max}^{3-\delta} ) \right]^{\frac{1}{1-\delta}}\ .
\end{eqnarray}
However, $R_{\rm ph}^\tau$ may be smaller than $R_{\rm min}$ at the
later time of evolution. We, therefore, define this photosphere
radius as $R_{\rm ph} = \max[R_{\rm ph}^\tau, R_{\rm
  min}]$. Eventually, the flux with frequency $\nu$ that is measured by the observer is obtained by summing up the contributions from all the rays
\begin{eqnarray}
  \label{eq:flux}
  F_{\nu} = \frac{2h\nu^3}{c^2} \int_{{\bm n}\cdot \bm{n}_{\Omega}>0} \frac{1}{\exp(h\nu/kT_{\rm eff})-1}
             \frac{R_{\rm ph}^2}{D^2} {\bm n}\cdot d\bm{\Omega} \ , \nonumber \\
\end{eqnarray}
where $\bm{n}$ and $\bm{n}_{\Omega}$ are the unit vector along the
line of sight, and the unit vector of the solid angle,
respectively. Subsequently, we determine the monochromatic AB magnitude by
$M_\nu = -2.5\log_{10}(F_\nu / 3631J)$.

Having determined the models for describing the EOS
and kilonovae, we will perform the Bayesian analysis of the EOS by exploiting the data of AT2017gfo~\citep{Villar2017}, GW170817~\citep{Abbott2019c}, and NICER pulsars~\citep{MCMiller2019b,
 Riley2019, Miller2021, Riley2021}. Before that, an introduction
of the details of Bayesian analysis will be presented in the next
section.

\begin{table}
\centering
   \vskip-3mm
   \caption{All of the parameters in the models and Bayesian
     analysis. Some of the parameters are not put into the
     analysis and sampling, and their prior distributions are denoted
     as Fixed. The $\log$ represent the uniform distribution of its
     logarithmic value, and the Constrained Uniform of the $\kappa$
     parameters denotes the uniform prior with the constrained
     condition $\kappa_{\rm high}>\kappa_{\rm low}$. }
   \setlength{\tabcolsep}{0.8pt} \renewcommand\arraystretch{1.2}
  \begin{ruledtabular}
    \begin{tabular}{lcccc}
      \multicolumn{5}{c}{The parameters and priors of EOS} \\ \hline
      Parameters & Unit &  Prior & Minimum & Maximum  \\ \hline
      $n_0$ & ${\rm fm}^{-3}$ & Fixed & $0.16$ & $0.16$ \\
      $E/A$ & MeV & Fixed   & $-16$ & $-16$ \\
      $J_0$ & MeV & Uniform & $30$  & $35$ \\
      $K_0$ & MeV & Uniform & $220$ & $280$ \\
      $L_0$ & MeV & Uniform & $20$  & $85$ \\
      $M_{\rm N}^\ast/M_{\rm N}$ & - & Uniform & $0.65$ & $0.80$ \\
    \end{tabular}
    \leavevmode\bigskip\\
    \begin{tabular}{lcccc}
      \multicolumn{5}{c}{The parameters and priors of ejecta and the binary} \\ \hline
      Parameters & Unit &  Prior & Minimum & Maximum  \\ \hline
      $\mathcal{M}$ & $\Msun$ & Uniform & $1.18$ & $1.21$ \\
      $q$ & - & Uniform & $1$  & $2$ \\
      $M_{\rm ej}$ & $\Msun$ & Uniform & $0.001$ & $0.01$ \\
      $\delta$    & - & Uniform & $1$ & $3$ \\
      $v_{\rm min}$ & $c$ & Uniform & $0.01$ & $0.15$ \\
      $v_{\rm max}$ & $c$ & Uniform & $0.18$ & $0.65$ \\
      $\kappa_{\rm low}$  & ${\rm cm\ g}^{-1}$ & Cons. Uniform & $0.1$ & $30$ \\
      $\kappa_{\rm high}$ & ${\rm cm\ g}^{-1}$ & Cons. Uniform & $0.1$ & $30$ \\
      $D$ & ${\rm Mpc}$ & Fixed & $40$ & $40$ \\
      $\theta_{\rm view}$ & Rad & Fixed & $\pi/6$ & $\pi/6$ \\
    \end{tabular}
    \leavevmode\bigskip\\
    \begin{tabular}{lcccc}
      \multicolumn{5}{c}{The central pressure of NICER sources} \\ \hline
      Parameters & Unit &  Prior & Minimum & Maximum  \\ \hline
      $p_{{\rm c}_1}$ & $10^{34}\ {\rm dyn/cm}^2$ & $\log$ & $2.774$ & $122.051$ \\
      $p_{\rm c}$    & $10^{34}\ {\rm dyn/cm}^2$ & $\log$ & $2.774$ & $122.051$ \\
    \end{tabular}
    \leavevmode\bigskip\\
    \begin{tabular}{lcccc}
      \multicolumn{5}{c}{Three additional parameters to denote the deviations } \\
      \multicolumn{5}{c}{of quasi-universal relations for modelling kilonova} \\ \hline
      Parameters & Unit &  Prior & Mean & Deviation  \\ \hline
      $\alpha_{m}$  & - & Gaussian & $0$ & $0.2$ \\
      $\alpha_{v}$  & - & Gaussian & $0$ & $0.2$ \\
      $\alpha_{e}$  & - & Gaussian & $0$ & $0.2$ \\
    \end{tabular}
  \end{ruledtabular}
  \label{tb:prior}
\end{table}

\section{Observational constraints and Bayesian analysis}
\label{sec:analyses}
Given a model hypothesis with a set of parameters ${\bm \theta}$, and
some data $d$, the posterior probability can be obtained by applying
the Bayes theorem,
\begin{eqnarray}
  \label{eq:bayes}
  p({\bm \theta}|d) = \frac{\mathcal{L}(d|{\bm \theta}) p({\bm \theta})}{\int
  \mathcal{L}(d|{\bm \theta}) p({\bm \theta}) d{\bm \theta}}\ ,
\end{eqnarray}
where $\mathcal{L}(d|{\bm \theta})$ denotes the likelihood of the data
$d$ given a set of parameters ${\bm \theta}$ and their corresponding
prior probability $p({\bm \theta})$. The denominator is the evidence
of data $d$ and acts as a normalization factor. The evidence can
be obtained by integrating the numerator all over the parameter
space. In reality, however, the parameter space has a non-trivial
number of dimensions, and may lead to a severe problem that is often
referred to as ``the curse of dimensionality". One can only resort to
the statistical computational techniques, e.g., Markov Chain Monte Carlo
or Nested Sampling methods, to approximate the evidence or the
marginalized distributions. In our analysis, the python package
\texttt{BILBY}~\citep{Ashton2019, Romero2020} and the nested sampler
\texttt{pymultinest}~\citep{Buchner2014} will be implemented to
generate the posterior samples and estimate the marginalized
distributions.

To incorporate the data of the kilonova light curve, the gravitational wave, and the NICER mass-radius measurements, we take the total likelihood function as the form of
\begin{eqnarray}
  \label{eq:bayes_t}
 \mathcal{L}(d|{\bm \theta}) = \mathcal{L}_{\rm AT2017gfo} \times \mathcal{L}_{\rm GW170817} \times \mathcal{L}_{\rm NICER} \ .
\end{eqnarray}
More details of the likelihood are described below in Sec.~\ref{subsec:likelihood}.

\subsection{Parameters and priors}
As explained above in Sec.~\ref{sec:eos}, the parameters of RMF models
can be directly related to the saturation properties of nuclear matter. 
Therefore, the six saturation properties will be treated as free parameters in our
Bayesian analysis. In practice, the first two properties, $n_0$
and $E/A$ are well determined and have much smaller uncertainties
compared with the rest four properties. We fix their value to be
$n_0=0.16\ {\rm fm}^{-3}$ and $E/A = 16$ MeV. The prior distribution
of the rest four properties are denoted as ${\bm \theta}_{\rm eos}$ and
are set as uniform distribution with the ranges displayed in
Table~\ref{tb:prior}

Furthermore, there are eight parameters in our kilonova model: 
The ejected mass $M_{\rm ej}$, the index of the mass distribution $\delta$, the
minimal and maximal velocity $v_{\rm min}$ and $v_{\rm max}$, the low
and high opacity value $\kappa_{\rm low}$ and $\kappa_{\rm high}$, the
luminosity distance of the source $D$ and the viewing angle
$\theta_{\rm view}$. 
In our analysis, the distance and viewing angle
are fixed as $D=40$ Mpc and $\theta_{\rm view}=\pi/6$. 
We denote the rest six parameters as ${\bm \theta}_{\rm kn}$ and compute their
posterior distribution in the following analysis.

The six input kilonova parameters ${\bm \theta}_{\rm kn}$ describing the properties of ejecta do relate to the binary parameters. Indeed, the quasi-universal relations are extracted by fitting the data of simulations~\citep[see][for more details]{Nedora2021, Nedora2022} and the ejected mass $M_{\rm ej}$, the mean velocity $v_{\rm mean}$ and the electron fraction $Y_{e}$ are expressed as functions of binary parameters (mass ratio $q$ and reduced tidal parameter $\tilde{\Lambda}$). However, these relations are not exact and deviations from the fitted formulations are expected in realistic situations. We introduce three deviation parameters to account for the uncertainty of the relations accordingly~\citep{Breschi2021}. Consequently, the $M_{\rm ej}$, $v_{\rm mean}$ and $Y_{e}$ are expressed with three additional deviation parameters $\alpha_m$, $\alpha_v$ and $\alpha_e$ as
\begin{eqnarray}
  \label{eq:deviations}
  \log_{10} M_{\rm ej} & = & (1 + \alpha_m) \log_{10} M_{\rm ej}^{\rm fit}(q, \tilde{\Lambda})\ , \\
  \label{eq:deviations2}
  v_{\rm mean} & = & (1 + \alpha_v)v_{\rm mean}^{\rm fit}(q, \tilde{\Lambda})\ , \\
  \label{eq:deviations3}
  Y_{e} & = & (1 + \alpha_e)Y_{e}^{\rm fit}(q, \tilde{\Lambda})\ .
\end{eqnarray}
The three deviation parameters ${\bm \theta}_{\rm dev}=(\alpha_m,\alpha_v, \alpha_e)$ will be treated
as input parameters in our Bayesian analysis, and their priors follow
the Gaussian distribution with vanished means and standard deviations
of $0.2$. 

In our kilonova models that take $v_{\rm min}$ and
$v_{\rm max}$ as the input parameters, the mean velocity can be
expressed in terms of the minimal and maximal velocity as
\begin{eqnarray}
  \label{eq:v_mean}
  v_{\rm mean} & = & \frac{(3-\delta)(v_{\rm max}^{4-\delta} - v_{\rm min}^{4-\delta})}
                     {(4-\delta)(v_{\rm max}^{3-\delta} - v_{\rm min}^{3-\delta})}\ .
\end{eqnarray}
The electron fraction $Y_{e}$ can be mapped into the mean opacity
$\bar{\kappa}$ of the ejecta by the relation
in~\citet{Tanaka2020}. The $\bar{\kappa}$ can be written as
\begin{eqnarray}
  \label{eq:kappa_mean}
  \bar{\kappa} = \frac{\sqrt{2}}{2}\kappa_{\rm high} + (1-\frac{\sqrt{2}}{2})\kappa_{\rm low}\ .
\end{eqnarray}
Meanwhile, once an EOS is determined from the EOS
parameters ${\bm \theta}$, the binary properties (mass ratio $q$ and
$\tilde{\Lambda}$) can be determined with given masses. Therefore, all
of the kilonova parameters ${\bm \theta}_{\rm kn}$ can be mapped into
the EOS parameters ${\bm \theta}_{\rm eos}$ with three deviation parameters
and two binary properties parameters (we use the mass ratio $q$ and the chirp mass
$\mathcal{M}$ in our analysis) by utilizing these above relations.

Finally, two additional parameters ${\bm \theta}_{\rm nicer}$ are
required for the NICER data, which represent the central pressure of
PSR J0030+0451 and PSR J0740+6620. All of the parameters and their
prior distributions are summarily displayed in Table~\ref{tb:prior}.

\begin{figure}
\vspace{-0.3cm}
{\centering
\includegraphics[width=0.49\textwidth]{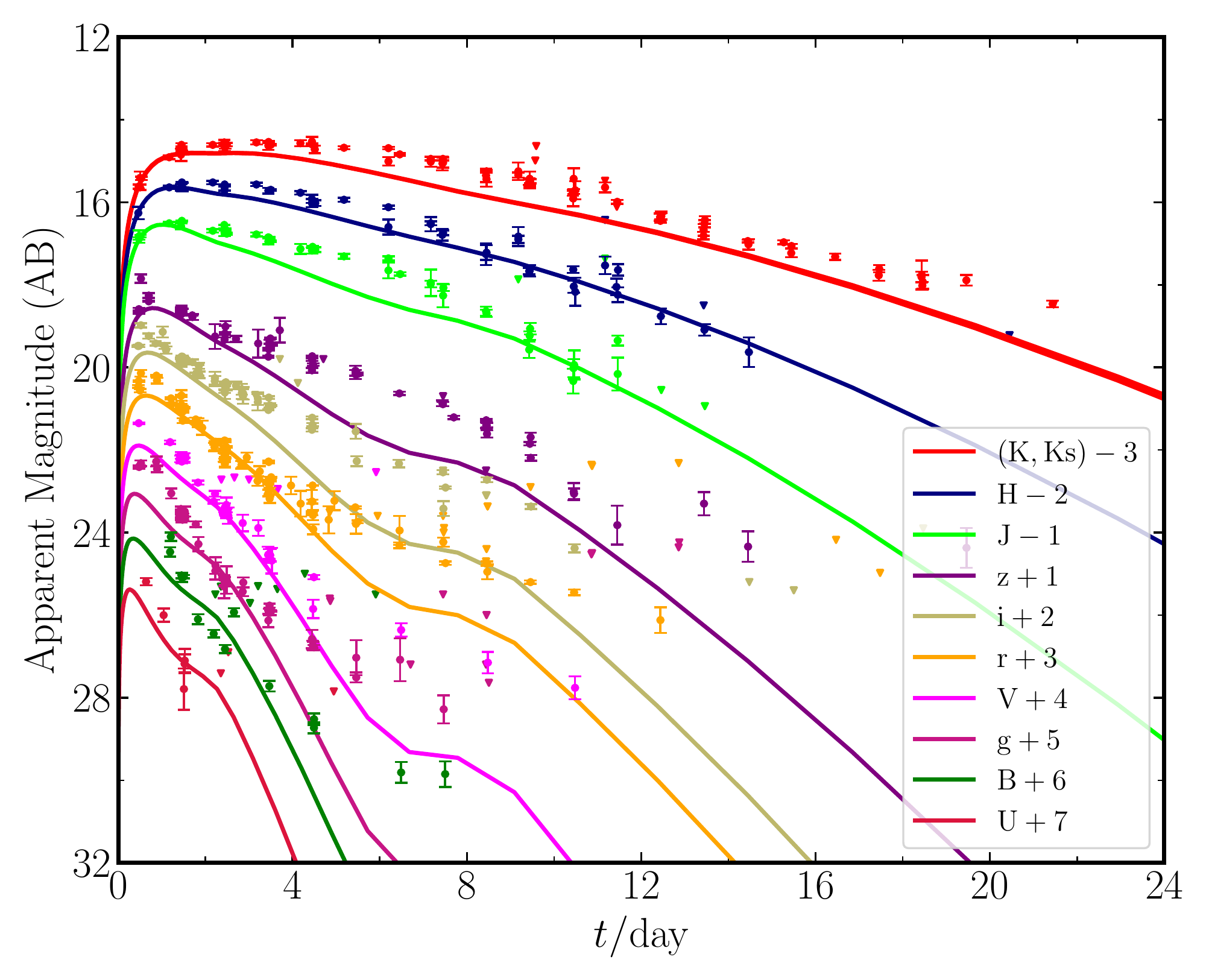}}
\caption{The light curves of the kilonova emission with the
  best-fitting parameters of ${\bm \theta}_{\rm kn}$, the solid lines
  with different color represent the predictions of the model of
  various bands. The observational data (circles) or limits
  (triangles) are taken from \citet{Villar2017}.}\label{fig:lc}
\vspace{-0.1cm}
\end{figure}

\begin{table}[t!]
\centering
\caption{The posterior results of the kilonova model parameters
  $\bm{\theta}_{\rm kn}$, the best fit values, the median value, and
  the $90\%$ confidence interval for each parameter are displayed.}
   \setlength{\tabcolsep}{0.8pt} \renewcommand\arraystretch{1.2}
  \begin{ruledtabular}
    \begin{tabular}{lcc}
      Parameters  &  Best fit & Median($90\%$)  \\ \hline
      $M_{\rm ej}$\ ($10^{-2}\Msun$) & $2.5977$ & $2.5972^{+0.0115}_{-0.0109}$ \\
      $\delta$    & $1.3078$ & $1.3081^{+0.0091}_{-0.0095}$ \\
      $v_{\rm min}($c$)$ & $0.1061$ & $0.1061^{+0.0004}_{-0.0004}$ \\
      $v_{\rm max}$($c$) & $0.5066$ & $0.5065^{+0.0037}_{-0.0036}$ \\
      $\kappa_{\rm low}$(${\rm cm\ g}^{-1}$)  & $0.7844$ & $0.7843^{+0.0044}_{-0.0041}$ \\
      $\kappa_{\rm high}$(${\rm cm\ g}^{-1}$) & $6.9509$ & $6.9487^{+0.0495}_{-0.0462}$ \\
    \end{tabular}
  \end{ruledtabular}
  \label{tab:kn}
\end{table}

\subsection{The observational data and likelihood}
\label{subsec:likelihood}

\subsubsection{AT2017gfo}
The observed light curves of AT2017gfo~\citep{Villar2017} will be fitted by our kilonova model. In reproducing the light curve of AT2017gfo, we only consider the dynamical ejecta as the source of $r$-process nucleosynthesis, since the effects of EOS on other parts of ejecta are mild~\citep[see \eg][]{Perego2017, Yu2018, Ren2019, Breschi2021}. 
In Fig.~\ref{fig:lc}, we plot the light curves of the kilonova model
with the best-fitting parameters of ${\bm \theta}_{\rm kn}$. The
observational data (circles) or limits (triangles) are taken
from~\citet{Villar2017}. Note that the solid lines that represent the
model predictions deviate from the observational data significantly after 4 days of the merger event for most of the bands (only K, H, J bands are compatible). This might be the consequence that only two components (red and blue) are taken into account in our model. 
As generally believed, 
a third component should be incorporated to account for it~\citep[e.g.,][]{Perego2017, Villar2017, Yu2018, Ren2019, Breschi2021,Qi2022},
including the energy or material injection from the central black hole hyperaccretion systems or magnetars, which is independent to the EOS. Therefore we will not address further in the following.

After obtaining the posterior samples of kilonova parameters
${\bm \theta}_{\rm kn}$, we approximate their posterior distribution
with the Gaussian kernel density estimation (KDE). In the following, the
posterior distribution will be treated as the likelihood of the EOS
parameters ${\bm \theta}_{\rm eos}$ and the deviation parameters
${\bm \theta}_{\rm dev}$.

\subsubsection{GW170817}
The GW170817 likelihood is calculated through a high-precision interpolation of the likelihood developed in~\citet{Hernandez2020} from fitting the strain data released by LIGO/Virgo,
which is encapsulated in the python package \texttt{toast},
\begin{eqnarray}
  \label{eq:llh_gw}
  \mathcal{L}_{\rm GW170817} = F(\Lambda_1, \Lambda_2, \mathcal{M}, q)\ ,
\end{eqnarray}
where the chirp mass is $\mathcal{M}=(M_1M_2)^{3/5}/(M_1+M_2)^{1/5}$, the mass ratio is $q=M_1/M_2$, and
$\Lambda_1$($M_1$) and $\Lambda_2$($M_2$) denote the tidal deformability (mass) of the individual star, respectively. $\Lambda_1$ and $\Lambda_2$ are connected with the reduced tidal deformability
$\tilde{\Lambda}$ by
\begin{eqnarray}
  \label{eq:t_lam}
  \tilde{\Lambda} = \frac{16}{13} \frac{(q+12)q^4 \Lambda_1 + (1+12q)\Lambda_2}{(1+q)^5}\ .
\end{eqnarray}
The tidal deformability, mass, and radius of a star can be computed by solving the perturbed tidal field equation~\citep{Flanagan2008, Hinderer08, Hinderer09} and the TOV equation simultaneously. Once the EOS and the central pressure of the star are determined, one can integrate both equations from the stellar center to the surface, where the pressure vanishes.

\subsubsection{PSR J0030+0451 and PSR J0740+6620}
The mass-radius measurements of two pulsars PSR J0030+0451 and PSR
J0740+6620 by NICER collaborations have set strong constraints on the EOS. 
At 68\% confidence level, the mass and
radius of PSR J0030+0451 are $M=1.34^{+0.15}_{-0.16}\Msun$,
$R=12.71^{+1.14}_{-1.19}$ km by~\citet{Riley2019}, or
$M=1.44^{+0.15}_{-0.14}\Msun$, $R=13.02^{+1.24}_{-1.06}$ km
by~\citet{MCMiller2019b}; and the results of PSR J0740+6620 are
$M=2.072^{+0.067}_{-0.066}\Msun$, $R=12.39^{+1.30}_{-0.98}$ km
by~\citet{Miller2021}, or $M=2.062^{+0.090}_{-0.091}\Msun$,
$R=13.71^{+2.61}_{-1.50}$ km by~\citet{Miller2021}. We implement
ST+PST model samples of PSR J0030+0451~\citep{Riley2019b} and the
NICER x XMM samples of PSR J0740+6620~\citep{Riley2021b} with the KDE
methods to generate the posterior distributions, which will be treated
as the likelihood in our analysis.
Note that the central pressure for these two pulsars is included and treated as input parameters when calculating the NICER likelihood. The masses and radii will be computed by solving the TOV equation with the EOS with given central pressures.

\begin{figure}
{\centering
  \includegraphics[width=0.48\textwidth]{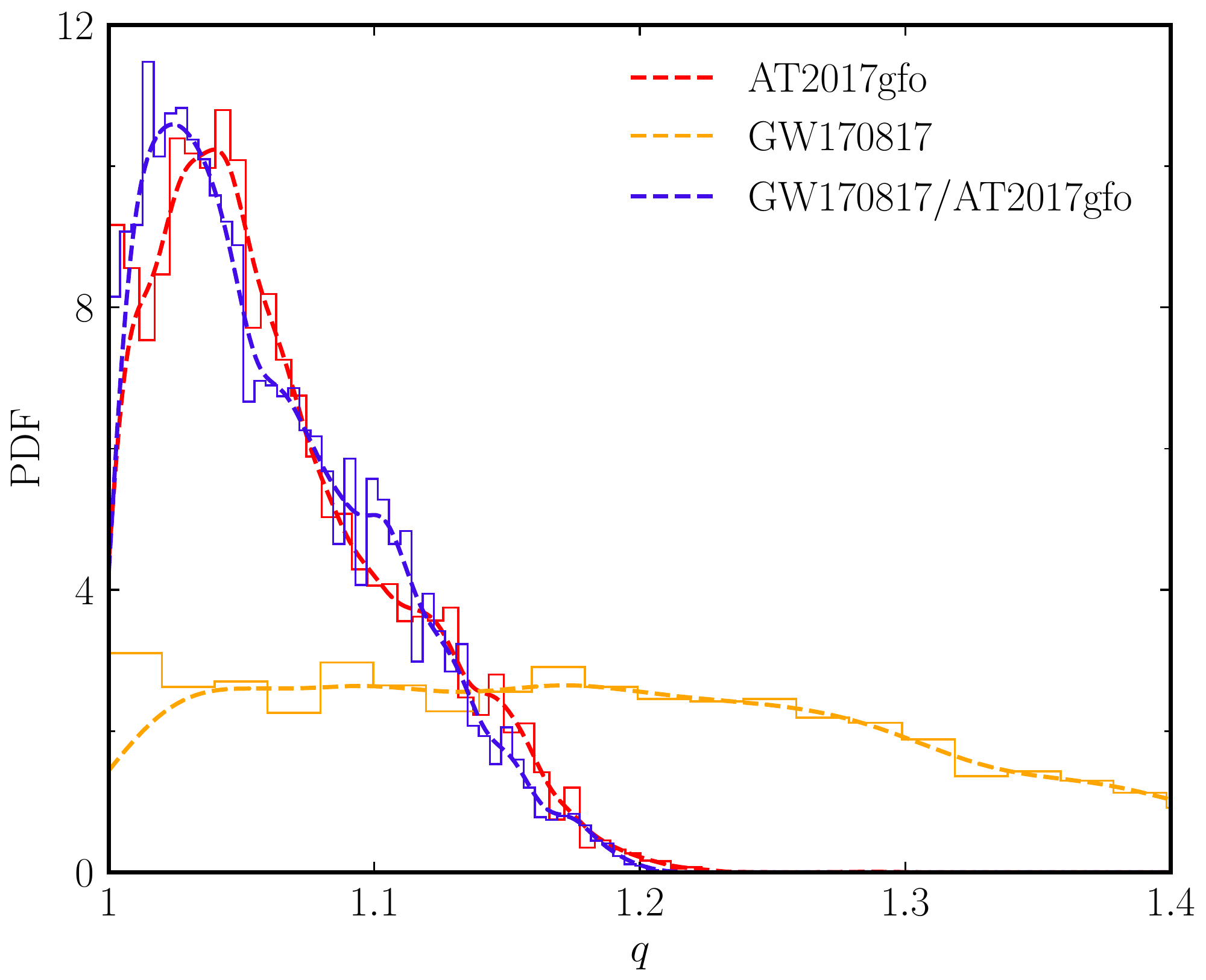}
  \includegraphics[width=0.48\textwidth]{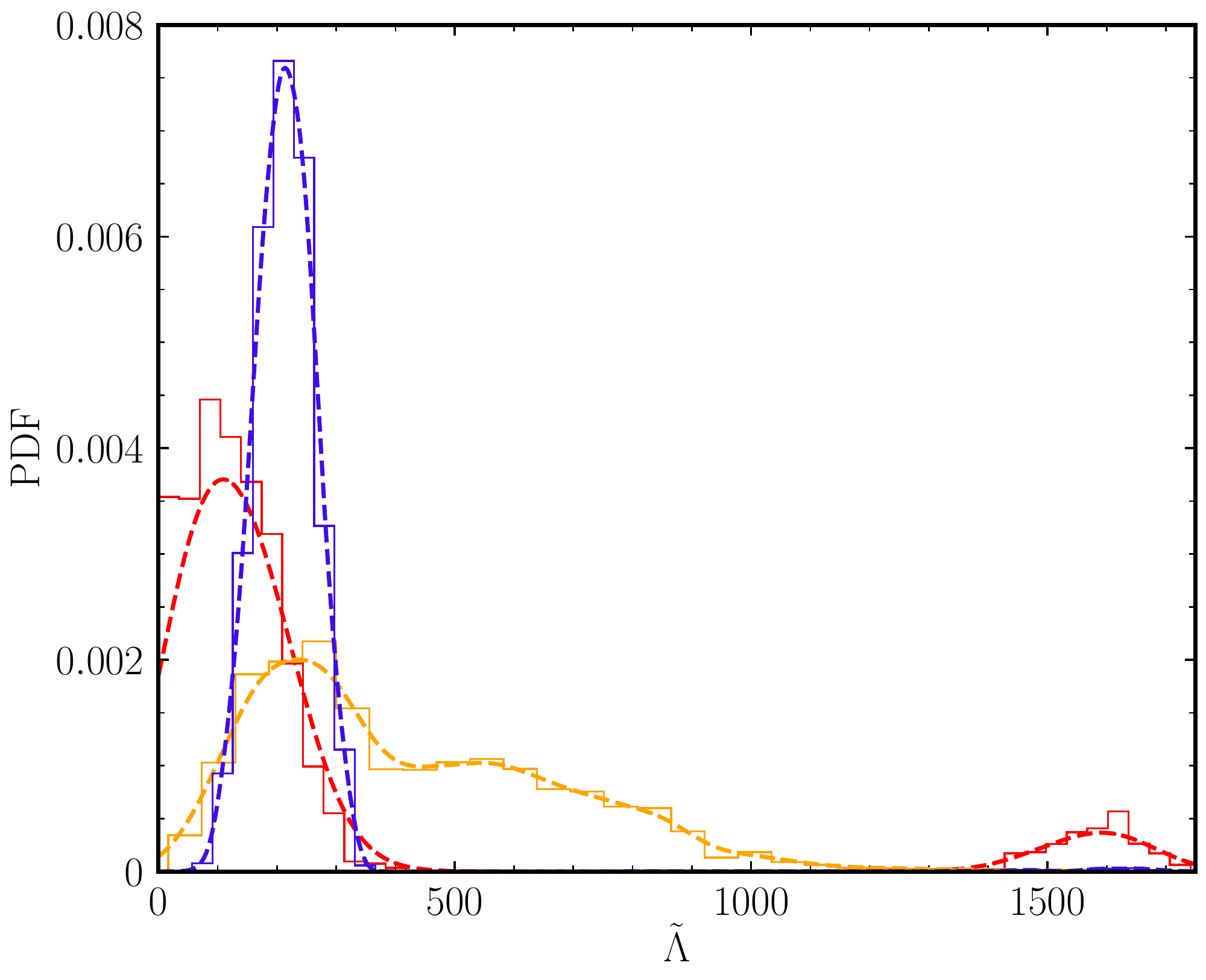}}
\caption{The posterior distributions of $q$ (upper) and
  $\tilde{\Lambda}$ (lower) for AT2017gfo (red), GW170817 (orange) and
  the AT2017gfo/GW170817 (blue) result. The solid lines represent
  the histogram of the posterior samples, and the dash lines are
  smoothed by the Gaussian KDE methods from the
  histogram. }\label{fig:qlam}
\vspace{-0.1cm}
\end{figure}

\begin{table}[t!]
\centering
\caption{The median value and $90\%$ confidence interval of $q$ and
  $\tilde{\Lambda}$ for different likelihood data. The large value of
  the confidence interval upper limit of $\tilde{\Lambda}$ for
  AT2017gfo is the result of the second peak
  in the posterior distributions.}
\setlength{\tabcolsep}{0.8pt} \renewcommand\arraystretch{1.2}
  \begin{ruledtabular}
    \begin{tabular}{lcc}
      Likelihood  & $q$ & $\tilde{\Lambda}$  \\ \hline
      AT2017gfo & $1.0530_{-0.0476}^{+0.0972}$ & $127.1565_{-112.5691}^{+1444.4892}$ \\
      GW170817  & $1.1871_{-0.1706}^{+0.2656}$ & $350.6256_{-243.0834}^{+530.0389}$ \\
      AT2017gfo/GW170817 & $1.0513_{-0.0452}^{+0.0916}$ & $213.5724_{-80.1237}^{+80.5000}$ \\
    \end{tabular}
  \end{ruledtabular}
  \label{tab:qlam}
\end{table}

\section{Results and Discussions}
\label{sec:results}
The EOS is connected with the ejecta properties through
the quasi-universal relations, which are the functions of the mass
ratio $q$ and the reduced tidal deformability $\tilde{\Lambda}$. In
our analysis, the parameters of kilonova model $\bm{\theta}_{\rm kn}$
will first be sampled. Their posterior results are displayed in Table~\ref{tab:kn}. 
The posterior distributions of the ejecta
parameters will be approximated by implementing the Gaussian KDE
method, and these posterior will further be used as the likelihood of
kilonova observations when sampling the binary parameters or EOS
parameters.

\begin{figure*}
{\centering
  \includegraphics[width=0.245\textwidth]{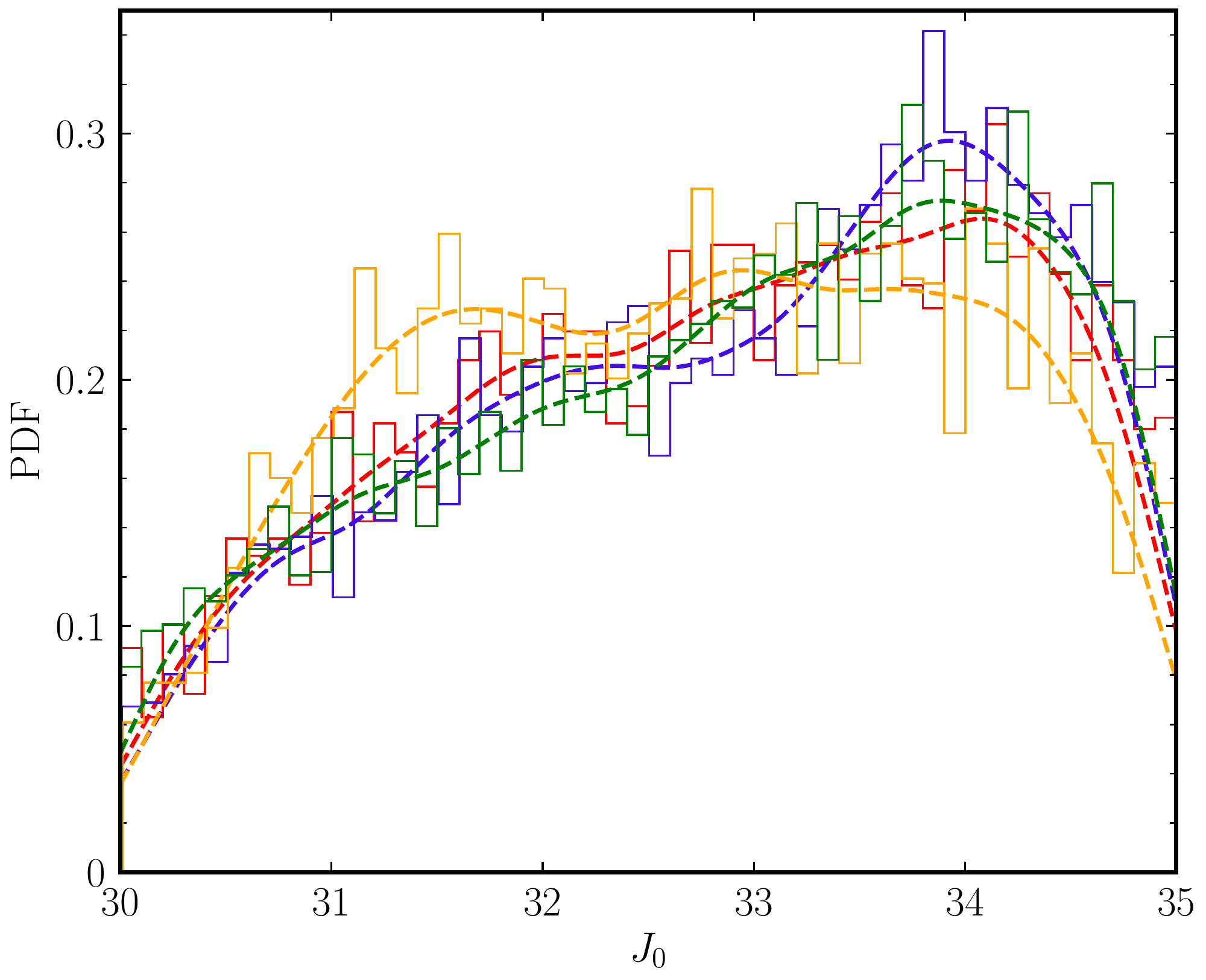}
  \includegraphics[width=0.245\textwidth]{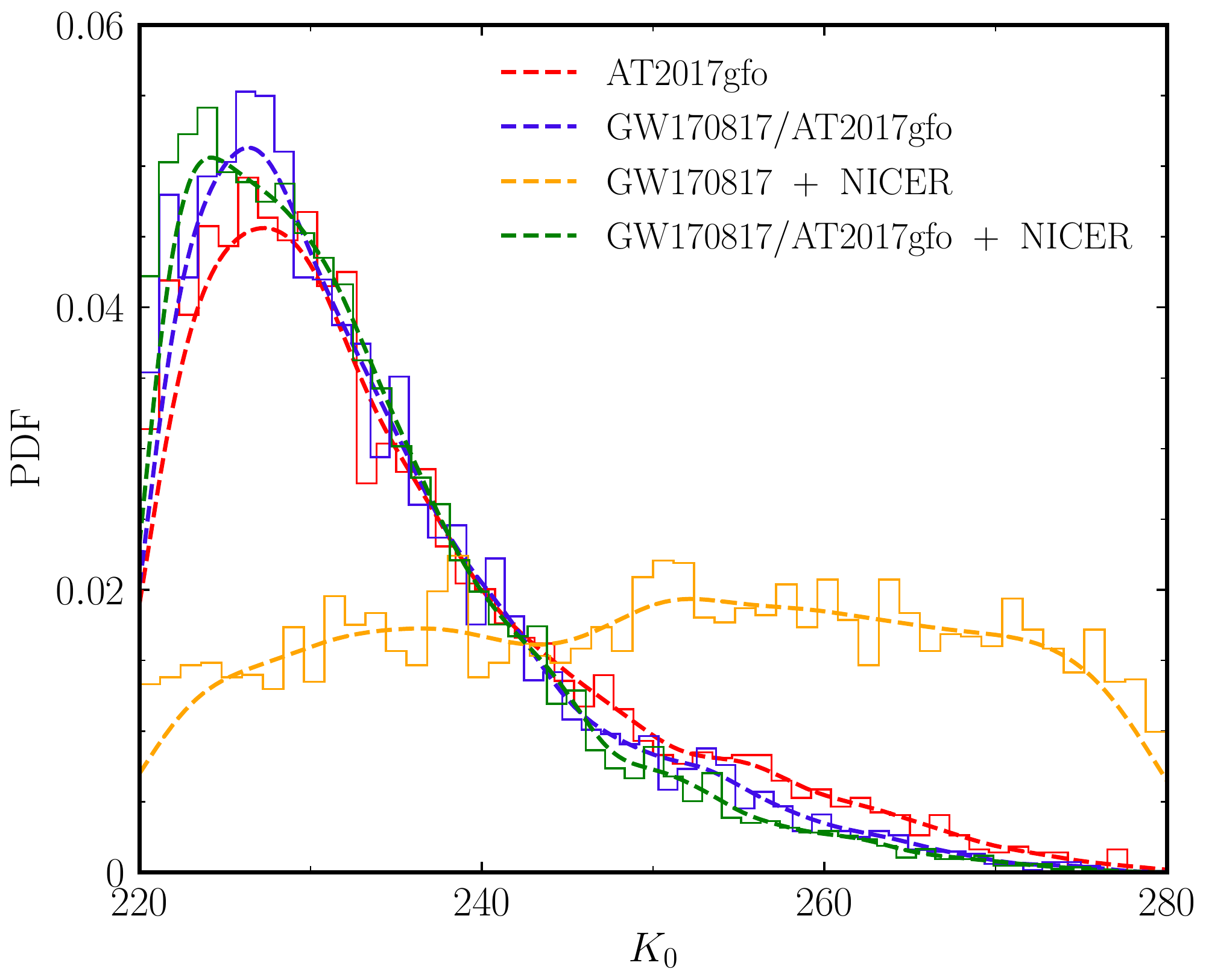}}
  \includegraphics[width=0.245\textwidth]{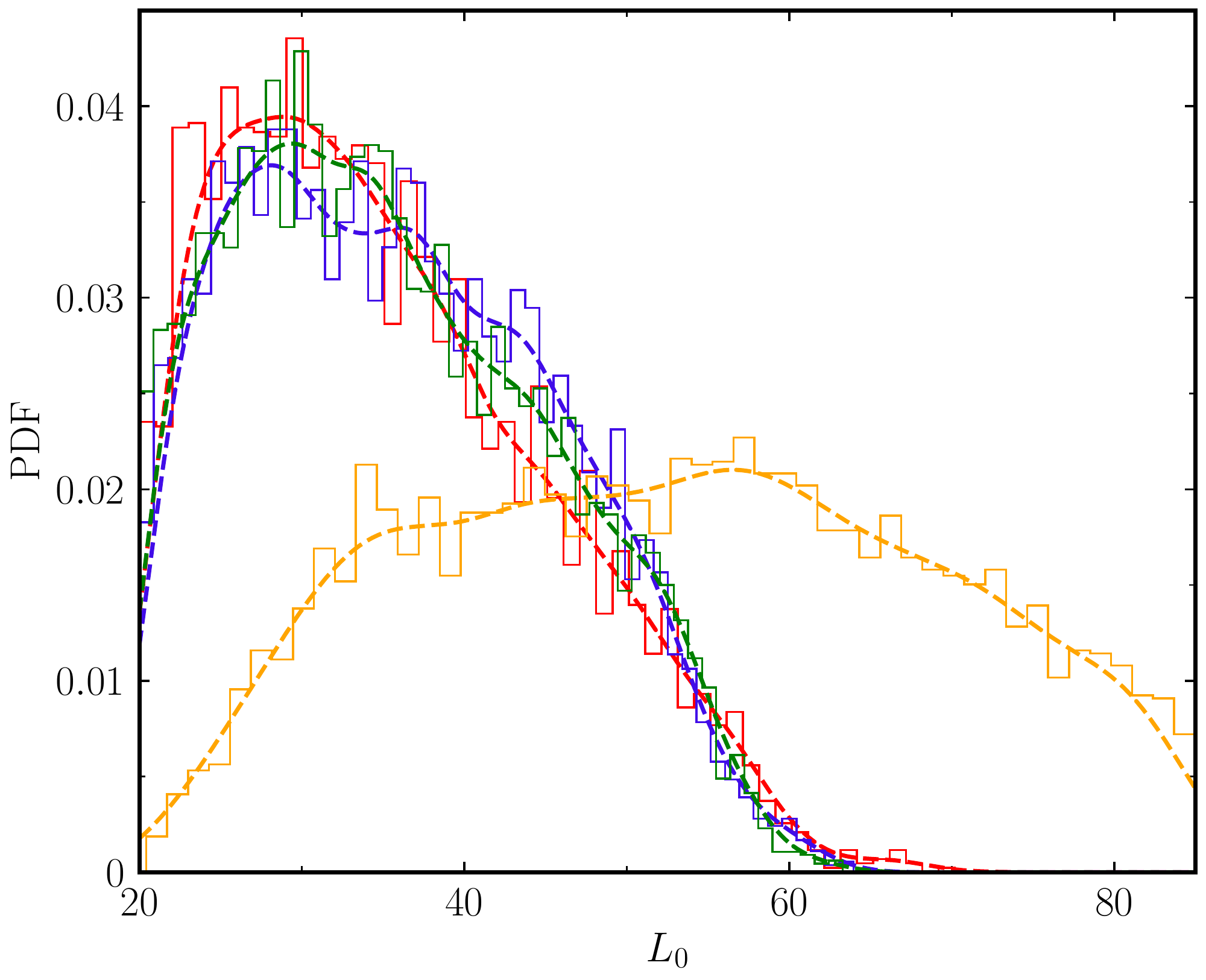}
  \includegraphics[width=0.245\textwidth]{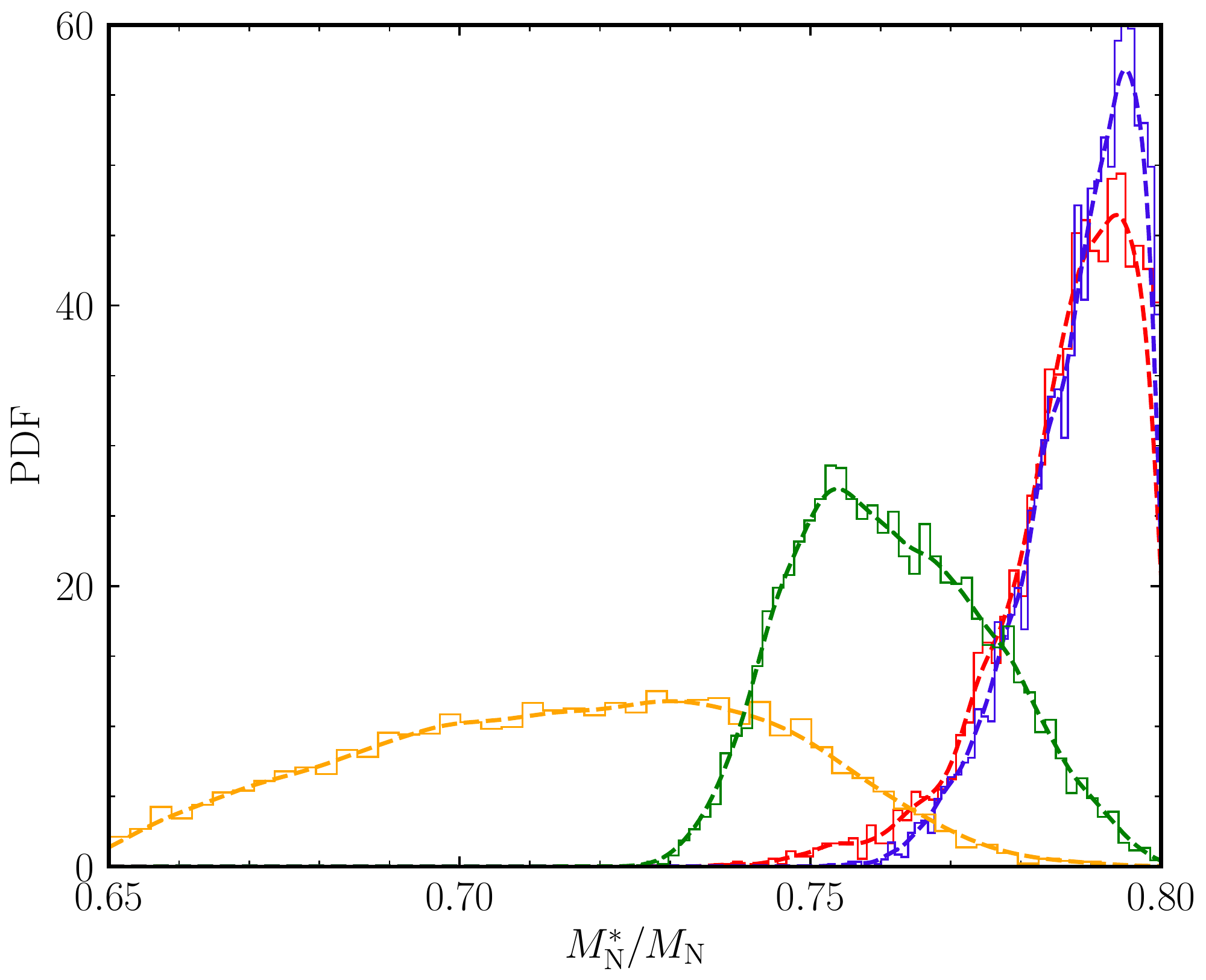}
  \caption{The posterior distributions of RMF EOS parameters
    $\bm{\theta}_{\rm eos}$. Be the same with Fig.~\ref{fig:qlam}, the
    histogram of the posterior samples and the smoothed distribution
    functions by the Gaussian KDE are represented by the solid and
    dash lines, respectively. The results from different analyses,
    AT2017gfo (red), GW170817/AT2017gfo (blue), GW170817 + NICER
    (orange) and GW170817/AT2017gfo + NICER (green), are denoted by
    different colors.  
    }\label{fig:sat}
  \vspace{-0.1cm}
\end{figure*}

\begin{table*}
\centering
   \vskip-3mm
   \caption{$90\%$ confidence intervals of the EOS parameters
     $\bm{\theta}_{\rm eos}$, or the saturation properties for nuclear
     matter, and the stellar properties constrained by four different analyses within the RMF framework.}   \vspace{-0.3cm}
   \setlength{\tabcolsep}{0.8pt} \renewcommand\arraystretch{1.2}
  \begin{ruledtabular}
  \begin{tabular}{lcccc}
   Parameters & AT2017gf &  GW170817/AT2017gf & GW170817+\nicer & GW170817/AT2017gf+\nicer  \\
    \hline 
    $J_0$ (MeV) & $32.9321_{-2.3807}^{+1.8031}$   & $33.0866_{-2.4922}^{+1.6679}$
                & $32.7278_{-2.1386}^{+1.9317}$   & $33.0410_{-2.5494}^{+1.7229}$ \\
    $K_0$ (MeV) & $231.5976_{-10.1760}^{+27.6173}$ & $230.5804_{-9.2084}^{+23.9437}$
                & $250.8818_{-27.2728}^{+25.1076}$ & $230.2890_{-9.0966}^{+22.0389}$ \\
    $L_0$ (MeV) & $33.7546_{-11.6658}^{+19.8140}$  & $35.3533_{-13.1968}^{+17.1443}$
                & $53.1642_{-24.9273}^{+26.3730}$  & $34.4599_{-12.5515}^{+18.2543}$ \\
    $M_{\rm N}^\ast/M_{\rm N}$  &  $0.7887_{-0.0211}^{+0.0100}$ & $0.7904_{-0.0172}^{+0.0083}$
                             &  $0.7166_{-0.0517}^{+0.0446}$ & $0.7604_{-0.0198}^{+0.0250}$ \\
    $R_{1.4}\ ({\rm km})$  &  $11.4107_{-0.2229}^{+0.2875}$ & $11.3930_{-0.2123}^{+0.2364}$
                          &  $12.3821_{-0.5639}^{+0.5311}$ & $11.6367_{-0.2312}^{+0.2121}$ \\
    $\Lambda_{1.4} $  &  $255.0494_{-26.4017}^{+41.1186}$ & $251.2908_{-25.4602}^{+32.4407}$
                     &  $440.8698_{-107.7413}^{+123.4322}$ & $300.2940_{-36.7643}^{+26.9738}$ \\
  \end{tabular}
  \end{ruledtabular}
  \label{tab:sat}
\end{table*}

\subsection{GW170817: mass ratio and tidal deformability}
The quasi-universal relations with their deviations
~(\ref{eq:deviations})--(\ref{eq:deviations3}) describe the ejecta
properties in terms of the binary parameters (the mass ratio $q$ and the reduced
tidal deformability $\tilde{\Lambda}$). 
We first compute the posterior samples of $q$ and $\tilde{\Lambda}$ by implementing the
nested sampler, and compare the results to those of AT2017gfo, GW170817 and
AT2017gfo/GW170817 in Fig.~\ref{fig:qlam}. We report in detail
the median values and the $90\%$ confidence intervals in
Table~\ref{tab:qlam}.

Fig.~\ref{fig:qlam} reports the posterior distributions of the mass ratio $q$ (upper panel) and reduced tidal deformability $\tilde{\Lambda}$ (lower panel) by fitting the AT2017gfo light curve
data, the GW170817 likelihood, and the combined data of kilonova and gravitational wave, and they are represented by the red, orange and blue lines, respectively. The solid lines represent the histogram of the samples and the dashed lines represent the distributions fitted by Gaussian KDE. Note that the KDE results of distributions deviate from the histogram when $q$ is close to $1$, for a stiff boundary are set at $q=1$ and the cases with $q<1$ do not exist. 
However, the Gaussian KDE function may extend to the region of $q<1$ and result in a decline close to $q=1$. 
In the upper panel of the distributions of mass ratio, it is seen that in comparison to the results from the GW170817 data, AT2017gfo favors a smaller mass ratio.
In the lower panel of the $\tilde{\Lambda}$ distributions, an interesting aspect of the kilonova data is reported.
The result of the kilonova fitting displays a bimodal structure~\citep{Breschi2021}: The first and the dominant peak locate around
$\tilde{\Lambda}=114$, while the secondary one is around
$\tilde{\Lambda}=1610$. Because of the second peak, the $90\%$
confidence interval upper limit of $\tilde{\Lambda}$ is considerably
larger than the GW170817 and GW170817/AT2017gfo results (shown in Table~\ref{tab:qlam}). 
Moreover, the first peak is close to that of GW170817 posterior distribution and results in a significant enhancement around $\tilde{\Lambda}=213$ region in the result of the combined data. Nevertheless, the secondary peak is suppressed by the GW170817 data and disappeared.
In spite of the consistency of the location of the dominant
peaks, GW170817 results show a longer tail with a larger value of
$\tilde{\Lambda}$. Consequently, the AT2017gfo data strongly favor a smaller tidal deformability and softer EOS, which will be shown in the following sections.

\subsection{The nuclear EOS parameters and neutron star properties} 

The EOS is specified by four parameters in our analyses, which are
the symmetry energy $J_0$, incompressibility $K_0$, symmetry energy
slope $L_0$ and effective mass ratio $M_{\rm N}^\ast / M_{\rm N}$. In
our process of sampling, we first calculate the coupling constants from
these saturations properties, and calculate by solving the equations
of motion~(\ref{eq:eqs_mot1})--(\ref{eq:eqs_mot3}) the neutron star
core EOS after adding the lepton contribution. 
We then join the core EOS with the usual BPS crust one~\citep{Baym71b}. 
The mass, radius and tidal deformability of neutron stars will be obtained with the whole stellar EOS and the likelihood of various
cases are yielded.

We report the posterior distributions of EOS parameters
$\bm{\theta}_{\rm eos}$ in Fig.~\ref{fig:sat}, and collect the median values and the $90\%$ confidence intervals in Table~\ref{tab:sat}. The results of four different analyses with the
data of AT2017gfo (red), GW170817/AT2017gfo (blue), GW170817 + NICER
(orange) and GW170817/AT2017gfo + NICER (green) are reported in
Fig.~\ref{fig:sat}. Same with Fig.~\ref{fig:qlam}, the
histograms are denoted by the solid lines and the approximated
distributions of KDE are denoted by dash lines.

The symmetry energy reported in Fig.~\ref{fig:sat} (leftmost
panel) from different analyses shows similar distributions. This
similarity can also be found for the confidence intervals in
Table~\ref{tab:sat}, and representing the insensitivity of symmetry
energy $J_0$ on these observational data. 
All of our analyses favor 
smaller incompressibility except for the GW170817 + NICER (orange) (see the
second panel of Fig.~\ref{fig:sat}). 
For example, the median value of $K_0$ is around $250$ MeV for the GW170817 + NICER case, while it is
 around $230$ MeV for the other three analyses. Such a deviation is a
consequence of the massive pulsar PSR J0740+6620 from NICER, which strongly favors a stiff EOS and 
hence a larger $K_0$. 
On the contrary, the dominate peaks in $\tilde{\Lambda}$ distribution (see
Fig.~\ref{fig:qlam}) of both GW170817 and AT2017gfo (red and blue) analyses imply a preference for soft EOS and a smaller radius for neutron stars. This difference between GW170817/AT2017gfo and NICER can also be found in the distributions of symmetry energy slope $L_0$ and
effective mass ratio $M_{\rm N}^\ast / M_{\rm N}$. The $L_0$ of
GW170817 + NICER analysis tends to be a larger value and implies a
larger radius of neutron star~\citep{Zhu2018}.
Similarly, a larger nucleon effective mass, which is preferred by the AT2017gfo and GW170817/AT2017gfo analyses results in a softer
EOS~\citep{Hornick2018}.
The introduction of the NICER observational data in our likelihood requires stiffer EOS and smaller effective mass. 
Therefore, the distribution of GW170817 + NICER (orange) favors a smaller effective mass, and the ratio decreases from $0.79$ of
GW170817 and AT2017gfo analyses to $0.72$ of GW170817 +
NICER. 
Finally, a trade-off of effective mass is achieved by the
GW170817/AT2017gfo + NICER analyses that balanced the soft EOS
preference of GW170817/AT2017gfo and stiff one of PSR J0740+6620 in
NICER data. 

We recall the recent laboratory PREX-II experiment that measured the neutron skin of $^{208}$Pb and implied the symmetry energy slope as $L_0 = 106 \pm 37$ MeV~\citep{Reed2021}. The large central value from the PREX-II measurement deviates significantly from our analysis of the observational data (about $34.4\mev$) as seen in Table~\ref{tab:sat}. 
Nevertheless, considering the large deviation of the $L_0$ distribution from PREX-II, different analyses from laboratory experiments and astrophysical data could be compatible with each other. For example, the joint analysis of PREX-II and the more recent CREX~\citep{Adhikari2022} suggests low symmetry energy slopes, i.e., $L_0=15.3^{+46.8}_{-41.5}$~\citep{Zhang2022}, which is similar to our present results.

\begin{figure}[t!]
\hspace{-0.5cm}
{\centering
{\includegraphics[width=0.48\textwidth]{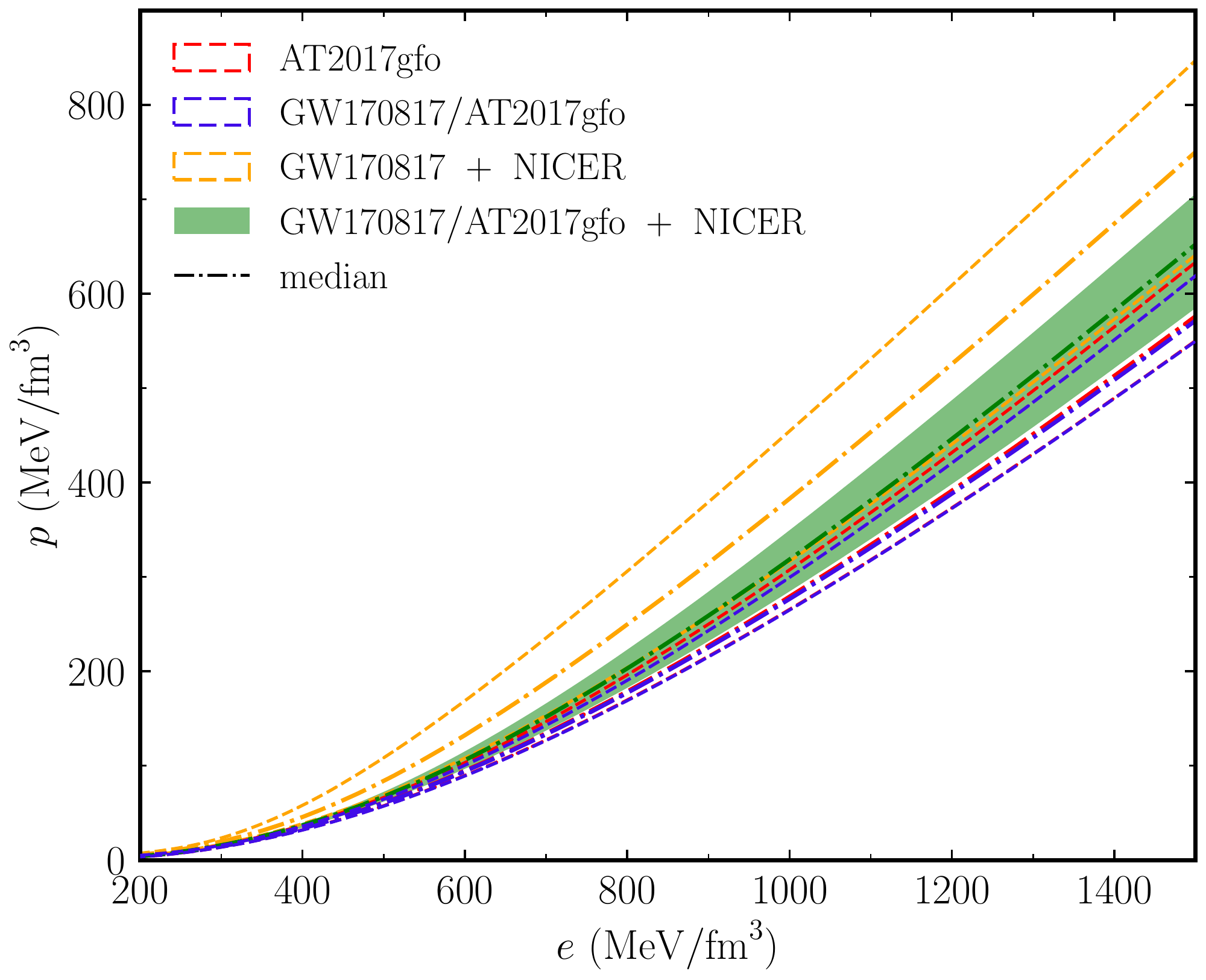}}
}
\caption{The $90\%$ confidence interval of EOS for all four
  analyses. The green shaded region represents the result of
  GW170817/AT2017gfo + NICER analysis, while other contours are
  represented by dash lines. The colors for each analyses are the same
  with the previous figure. The dash-dot lines denote the median
  results of the posterior distributions.  
  }\label{fig:eos}
\end{figure}

\begin{figure}[t!]
{\centering
{\includegraphics[width=0.48\textwidth]{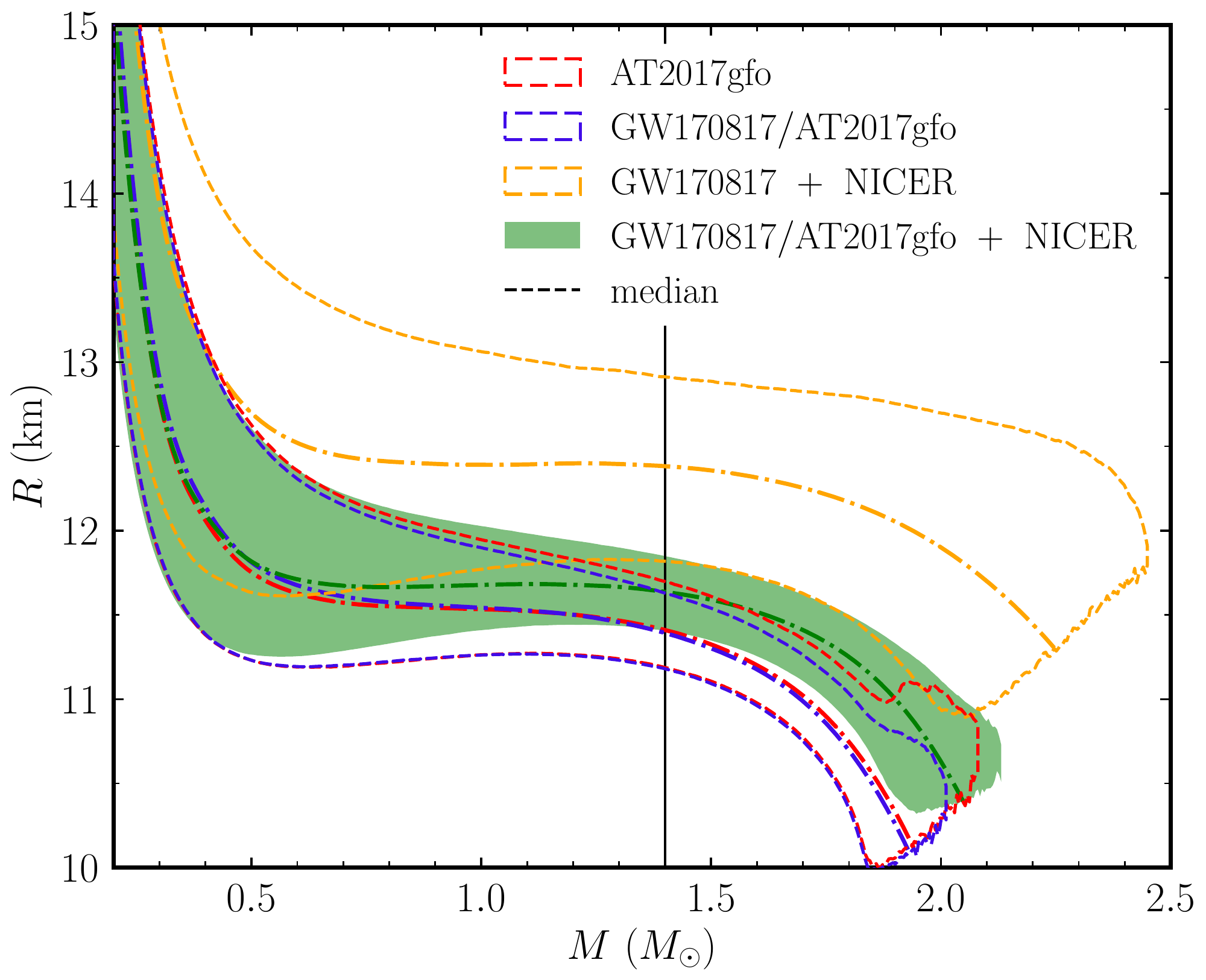}}
{\includegraphics[width=0.48\textwidth]{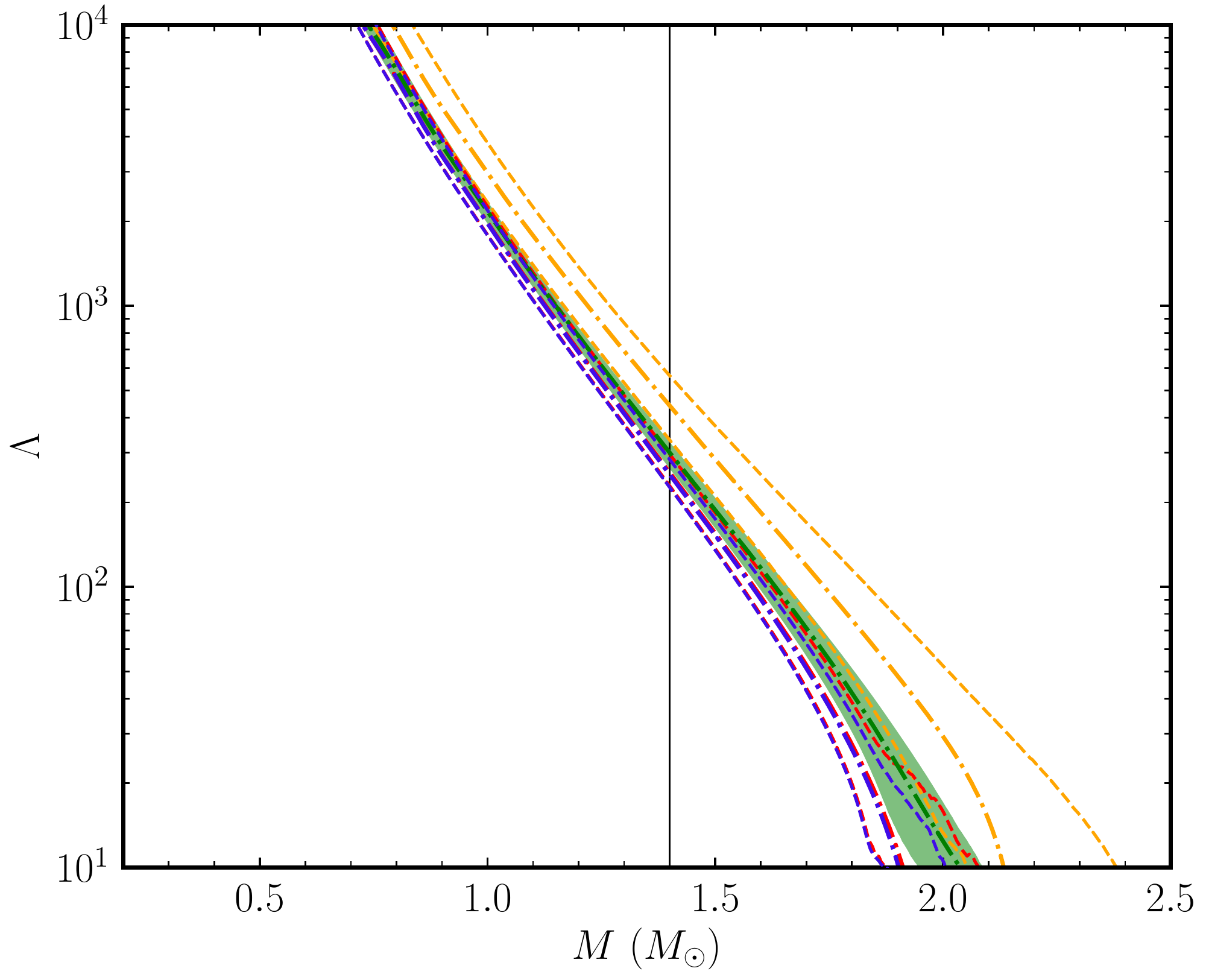}}
}
\caption{Same with Fig.~\ref{fig:eos}, but for the radius (upper
  panel) and tidal deformability (lower panel) as functions of the
  stellar mass. The vertical black line is the $1.4\Msun$
  line.  
  }\label{fig:mrtd}
\end{figure}

Fig.~\ref{fig:eos} compares and contrasts the $90\%$ confidence
intervals of the EOSs for all the analyses. The contour of
GW170817/AT2017gfo + NICER analysis is denoted by the green shaded
region, while the other contours are denoted by the dash lines. The
median results for each posterior distribution are denoted by the
dash-dot lines with the corresponding colors.
The results of the EOS confidence interval are consistent with the
saturation properties distributions and our previous discussions. 
Note that the secondary peak in $\tilde{\Lambda}$ disappeared in the distribution of EOS and saturation properties for AT2017gfo analysis. The large $\tilde{\Lambda}$ value of the secondary peak implies an unrealistically stiff EOSs, 
which are disfavored by the nuclear experimental results~\citep[e.g.,][]{Drischler2020, Zhang2021}. 
For example, in the current framework of the RMF model, an EOS with $E_{\rm sym}=43$ MeV, $K=300$ MeV $L=142$ MeV, $M_{\rm N}^\ast/M_{\rm N}=0.55$ could result in $\tilde{\Lambda}=1450$.

Moreover, the interval contours and the median lines of AT2017gfo and
GW170817/AT2017gfo are almost overlapping with each other. Reminding
the proximity of the dominate peak of $\tilde{\Lambda}$ distributions
from AT2017gfo and GW170817 analyses in Fig.~\ref{fig:qlam}, this
similarity on EOS is the result of that, and implies the consistency
of AT2017gfo and GW170817 data. 
On the other hand, the
analyses of the NICER data favor stiffer EOS because of the
massive pulsar. The medium region that fulfills the small
$\tilde{\Lambda}$ and large maximum mass $M_{\rm TOV}$ is
significantly enhanced in the distribution of the analysis that takes
all data into account. Meanwhile, the very soft and very stiff EOS is
disfavored.
Note that the upper bound of $M_{\rm TOV}$ in our analyses is around $2.1 \Msun$, to the 90\% posterior credible level, which is incompatible with GW190814~\citep{Abbott2020} if its $\sim 2.6\Msun$ low-mass component is assumed as a neutron star without phase transitions~\citep[see also discussions in][]{Li2021,Nathanail2021}. And the tension could in principle be resolved in the two-family scenario, which interprets the $\sim2.6\Msun$ component of GW190814 as a quark star and the GW170817 event as binary neutron star merger~\citep{Bombaci2021}.

Finally, we report the mass-radius relations and the tidal
deformability intervals of each analysis in Fig.~\ref{fig:mrtd}, and
display the radius and tidal deformability of $1.4\Msun$ stars
$R_{1.4}$ and $\Lambda_{1.4}$ in the last two rows of Table~\ref{tab:sat}, respectively. 
The analyses with GW170817 and AT2017gfo give 
smaller radius for stars around $1.4\Msun$ compared with the analyses with NICER data. 
The median value of $R_{1.4}$ increases from
$\sim 11.4$ km for AT2017gfo and GW170817/AT2017gfo
to $12.4$ km for GW170817 + NICER, and further decreases
to $11.6$ km for GW170817/AT2017gfo + NICER because of the
trade-off. 
We mention here that the radius results are similar to the ones obtained with a chiral effective-field-theory description of nuclear matter~\citep{Capano2020}.
The tidal deformability $\Lambda_{1.4}$ has the similar
behavior (increases from $\sim 250$ to $440$ and further goes down to
$300$) due to the positive correlation between $\Lambda_{1.4}$ and
$R_{1.4}$~\citep[e.g.,][]{Lim2018}.

\section{Conclusions}
\label{sec:sum}
Even since the first detection of the multimessenger signal of the GW170817 binary neutron star merger,
a large number of works have investigated its implications on the neutron
star EOS. The matter effects of the binary system imprinted into the
gravitational wave signal as the tidal deformability contributions,
and one may extract from it and constrain the EOS by analyzing the
the GW signals. 
On the other hand, the transient kilonova event of AT2017g
can also shed light on the neutron star EOS through the properties of
dynamical ejecta.  

In this work, we implemented the quasi-universal relations between the binary properties (the mass ratio $q$ and reduced tidal deformability $\tilde{\Lambda}$) and the ejecta properties (the ejected mass, velocity, and electron fraction), and combined the observational data of AT2017gfo to constrain the neutron star EOS. The reduced tidal deformability of binary can be directly related to the saturation properties of nuclear matter in the framework of the RMF model. Thereafter, we performed the Bayesian analysis of the EOS and the saturation properties (the symmetry energy $J_0$, incompressibility $K_0$, symmetry energy slope $L_0$, and effective mass ratio $M_{\rm N}^\ast / M_{\rm N}$) with the AT2017gfo light curve data. Our analysis shows a bimodal structure of the $\tilde{\Lambda}$ distribution, where the dominant peak corresponds to softer EOS and a smaller radius of stars. This dominant peak is enhanced by the GW170817 results, while the second peak is suppressed and disappeared in the distribution of GW170817/AT2017gfo. 

We proceed to perform joint analyses with various observational data combinations (GW170817/AT2017gfo, GW170817 + NICER, and GW170817/AT2017gfo + NICER). 
The $90\%$ confidence interval of EOS of AT2017gfo and GW170817/AT2017gfo were almost overlapping with each other, implying the consistency of GW170817 and AT2017gfo. However, the introduction of NICER data makes the posterior distributions strongly favor stiff EOS with a larger stellar radius, since the massive pulsar PSR J0740+6620 in the NICER data demands stiff EOSs to be consistent with it. 
As a result, both the very stiff and very soft EOSs are excluded for their incapability to reproduce the AT2017gfo data or PSR J0740+6620 data.
When combining all observational data, the nuclear matter properties at saturation are found to be  $J_0=33.0410_{-2.5494}^{+1.7229}\mev$, $K_0=230.2890_{-9.0966}^{+22.0389}\mev$, $L_0=34.4599_{-12.5515}^{+18.2543}\mev$ and $M_{\rm N}^\ast/M_{\rm N}=0.7604_{-0.0198}^{+0.0250}$, at 90\% confidence level. 
Correspondingly, the radius and the tidal deformability for $1.4\Msun$ neutron stars are
$11.6367_{-0.2312}^{+0.2121}$ km and $300.2940_{-36.7643}^{+26.9738}$, respectively. More future joint multimessenger observations on neutron stars, binary evolution, and their mergers are expected to further constrain their EOS.

\appendix
\section{Detailed derivation of nuclear matter properties from RMF model parameters}\label{A1}

We sum up the energy density and pressure expressions~(\ref{eq:ener}) -- (\ref{eq:press}) at saturation density ($p=0$) to yield a simplified expression:
\begin{eqnarray}
  \label{eq:eplusp}
  e + p = (E/A + M_{\rm N})n_0 = \sum_{i=n,p} e_{\rm kin}^i + \sum_{i=n,p} p_{\rm kin}^i + g_\omega \omega n_0\ .
\end{eqnarray}
The left-hand-side is known from $E/A$ and $n_0$, and the kinetic terms only depends on fermi momentum $p_{\rm F}$ and effective mass. Combine this equation with the $\omega$ equation of motion, we can express $\omega$ and $g_\omega$ in terms of the known 
quantities
\begin{eqnarray}
  \label{eq:omega}
  \omega  & = & \sqrt{\frac{(E/A + M_{\rm N}) n_0 - \sum_i (e_{\rm kin} + p_{\rm kin})}{m_\omega^2}}\ , \\
  \label{eq:gomega}
  g_\omega & = & \frac{m_\omega^2 \omega}{n_0}\ .
\end{eqnarray}
The expressions of $\Lambda_v$ and $g_\rho$ ($\rho$ vanishes for
symmetric nuclear matter) can also be obtained in the same way by
combining eqs.~(\ref{eq:eqs_mot3}), (\ref{eq:esym}) and (\ref{eq:l0})
\begin{eqnarray}
  \label{eq:lambda_c}
  \Lambda_v & = & - \frac{m_\omega^2 \alpha}{3 \beta^2 g_\omega^3 \omega n_0^2}\ , \\
  \label{eq:grho}
  g_\rho^q & = & \sqrt{\frac{m_\rho^2}{\beta^{-1} - \Lambda_v (g_\omega \omega)^2}}\ .
\end{eqnarray}
where $\alpha$ and $\beta$ is written as
\begin{eqnarray}
  \label{eq:beta}
  \alpha & = & L_0 - 3J_0 -  \frac{1}{2}\left(\frac{3\pi^2}{2}n_0 \right)^{2/3} \frac{1}{E_{\rm F}} \times \nonumber \\
         & & \left(\frac{g_\omega^2}{m_\omega^2 + \Lambda_v (g_\omega g_\rho \rho)^2}\frac{n_0}{E_{\rm F}}
             - \frac{K_0}{9E_{\rm F}} - \frac{1}{3} \right)\ , \\
  \beta & = & \frac{2J_0}{n_0} - \frac{p_{\rm F}^2}{3E_{\rm F} n_0}\ .
\end{eqnarray}

The last three parameters determination rely on eqs.~(\ref{eq:press}),
(\ref{eq:eqs_mot1}) and the derivative of (\ref{eq:eqs_mot1}). Their
expressions can be written as
\begin{eqnarray}
  \label{eq:sigma}
  \sigma & = & \sqrt{\frac{C - 6B + 12A}{m_\sigma^2}}\ , \\
  g_2 & = & \frac{-3C + 15B - 24A}{\sigma^3}\ , \\
  g_3 & = & \frac{2C - 8B + 12A}{\sigma^4}\ , \\
  g_\sigma & = & \frac{(g_\sigma \sigma)}{\sigma} \ ,
\end{eqnarray}
where $A$, $B$ and $C$ are
\begin{eqnarray}
  \label{eq:abc}
  A & = & \sum_{i=n,p} p_{\rm kin}^i + \frac{1}{2} m_\omega^2 \omega^2\ ,  \\
  B & = & (g_\sigma \sigma) n_s\ , \\
  C & = & -(g_\sigma \sigma)^2 \left[\frac{\partial n_{\rm s}}{d M_{\rm N}^\ast}
          + \frac{\partial n_{\rm s}}{\partial p_{\rm F}} / \frac{\partial M_{\rm N}^\ast}{dp_{\rm F}} \right]\ .
\end{eqnarray}
In these expressions, $(g_\sigma \sigma)$ can be evaluated by
$g_\sigma \sigma = M_{\rm N} - M_{\rm N}^\ast$,
$\partial M_{\rm N}^\ast / dp_{\rm F}$ is obtained from $K_0$.

\section*{Acknowledgements}
We are thankful to Jinping Zhu, Yanqing Qi, Enping Zhou and the XMU neutron star group for the helpful input and discussions. This work was supported by National SKA Program of China (No.~2020SKA0120300), the National Natural Science Foundation of China (Grant Nos.~11873040, 12273028, 12103033, 12173031 and 12221003), the Youth Innovation Fund of Xiamen (No.~3502Z20206061) and the China national postdoctoral program for innovation talents (No.~BX20220207).


\bibliographystyle{aasjournal}
\bibliography{EoS_kn.bib}

\end{document}